\documentclass{pas}

\usepackage{multirow}
\usepackage{multicol}
\usepackage{booktabs}
\newcommand{\ttrule}{\specialrule{0.8pt}{0pt}{0pt}
}

\newcommand*\degr{\ensuremath{^\circ}}
\newcommand*\arcmin{\ensuremath{^\prime}}
\newcommand*\arcsec{\ensuremath{^{\prime\prime}}}

\newcommand*\farcs{\ensuremath{\overset{\prime\prime}{.}}}

\begin{document}

\lefttitle{Preparation of a celestial field for precise VLBI observations of the JUICE spacecraft: a test case of the Venus flyby}
\righttitle{K. Perger et al.}

\jnlPage{1}{12}
\jnlDoiYr{2026}
\doival{10.1017/pasa.xxxx.xx}

\articletitt{Research Paper}

\title{Preparation of a celestial field for precise VLBI observations of the JUICE spacecraft: a test case of the Venus flyby}

\author{\gn{Krisztina} \sn{Perger}$^{1}$,
  \gn{S\'andor}~\sn{Frey}$^{1,2}$,
  \gn{Giuseppe}~\sn{Cim\`{o}}$^{3}$,
  \gn{Dominic}~\sn{Dirkx}$^{4}$,
  \gn{Jasper}~\sn{Edwards}$^{5,6}$,
  \gn{Sam}~\sn{Fayolle}$^{7,4}$,
  \gn{Judit}~\sn{Fogasy}$^{1}$,
  \gn{Krisztina~\'E.}~\sn{Gab\'anyi}$^{2,8,1}$,
  \gn{Leonid~I.}~\sn{Gurvits}$^{3,4}$,
  \gn{M\'{a}t\'{e}}~\sn{Krezinger}$^{1}$,
  \gn{Guifr\'{e}}~\sn{Molera~Calv\'{e}s}$^{5}$,
  \gn{Vidhya}~\sn{Pallichadath}$^{4}$,
  \gn{Noor~Masdiana~Md}~\sn{Said}$^{3}$,
  and \gn{Bert}~\sn{Vermeersen}$^{4}$
}

\affil{$^1$Konkoly Observatory, HUN-REN Research Centre for Astronomy and Earth Sciences, MTA Centre of Excellence, Konkoly Thege Mikl\'{o}s \'{u}t 15-17, 1121 Budapest, Hungary
$^2$Department of Astronomy, Institute of Physics and Astronomy, ELTE E\"otv\"os Lor\'and University, P\'azm\'any P\'eter s\'et\'any 1/A, 1117 Budapest, Hungary
$^3$Joint Institute for VLBI ERIC, Oude Hoogeveensedjik 4, 7991 PD Dwingeloo, The Netherlands
$^4$Faculty of Aerospace Engineering, Delft University of Technology, Kluyverweg 1, 2629 HS Delft, The Netherlands
$^5$School of Natural Sciences, University of Tasmania, Hobart, Tasmania, Australia
$^6$CSIRO Space and Astronomy, Epping, NSW, Australia
$^7$European Space Research and Technology Centre, ESA, Keplerlaan 1, 2201 AZ Noordwijk, The Netherlands
$^8$HUN-REN–ELTE Extragalactic Astrophysics Research Group, ELTE E\"otv\"os Lor\'and University, P\'azm\'any P\'eter s\'et\'any 1/A, 1117 Budapest, Hungary
}

\corresp{G. Molera Calv\'{e}s, Email: guifre.moleracalves@utas.edu.au}

\citeauth{Perger K, Frey S, Cim\'{o} G, Dirkx D, Edwards J, Fayolle S, Fogasy J, Gab\'{a}nyi K, Gurvits L, Krezinger M, Molera Calv\'{e}s G, Pallichadath V, Said N and Vermeersen B, High-resolution radio observations of potential phase-reference calibrator sources for the JUICE spacecraft during its Venus flyby. {\it Publications of the Astronomical Society of Australia} {\bf 00}, 1--12. https://doi.org/10.1017/pasa.xxxx.xx}

\history{(Received xx xx xxxx; revised xx xx xxxx; accepted xx xx xxxx)}

\begin{abstract}
 The Jupiter Icy Moons Explorer (JUICE) mission, launched in 2023, is on its way to the Jovian system, to arrive in 2031. The Planetary Radio Interferometry and Doppler Experiment (PRIDE) performs radio astronomical observations of the spacecraft already in its interplanetary cruise phase. Since the sky density of potential compact radio sources suitable as phase-reference calibrators in very long baseline interferometry (VLBI) observations is low, densifying the celestial reference frame along the Ecliptic is an essential task in support of JUICE.  In preparation for the JUICE gravity assist manoeuvre at Venus in August 2025, to support the accurate reconstruction of the spacecraft state vector, we selected in-beam calibrator source candidates along the JUICE trajectory. Additionally, we attempted to link known nearby bright compact reference sources from the Radio Fundamental Catalog (RFC) to the International Celestial Reference Frame (ICRF).   We performed nodding-style phase-referencing VLBI observations with the European VLBI Network (EVN) at $8.4$~GHz, targeting nine in-beam calibrator candidate (ICC) and three RFC sources in the celestial vicinity of JUICE's trajectory during its Venus flyby. We linked their positions to nearby ICRF sources by means of relative astrometry.  Four of the ICC targets and all RFC sources were detected and imaged with the EVN, and their coordinates determined with typically sub-milliarcsecond (mas) precision.  Although none of the targeted ICC sources proved suitable as VLBI phase-reference calibrators because of being too faint or resolved, their newly-determined positions and flux densities could in principle be used to verify the detection of the spacecraft at the time of the actual flyby around Venus. During the detailed analysis of the ICC targets, we found that the structure of one of them, J0843+1826 is resolved on mas scale, and the other one, J0843+1814 is a young radio galaxy, likely a member of a cluster.
\end{abstract}


\maketitle

\section{Introduction}

The European Space Agency (ESA) launched its interplanetary spacecraft, Jupiter Icy Moons Explorer \citep[JUICE,][]{{2023SSRv..219...53F}} from the Guiana Space Centre in Kourou, French Guiana on 14 April 2023. JUICE is expected to reach the Jovian system in July 2031. Its interplanetary trajectory in the inner Solar System \citep{2024SSRv..220...67B} is shaped by four gravity assist manoeuvres, a flyby at the Earth--Moon system (August 2024), at Venus (August 2025), and two more times around the Earth (September 2026 and January 2029). The JUICE science mission consist of $11$ experiments \citep{{2023SSRv..219...53F}}. Among them, the only one without dedicated on-board instrumentation is the Planetary Radio Interferometry and Doppler Experiment \citep[PRIDE,][]{2023SSRv..219...79G}. PRIDE is a multi-purpose experiment with its primary objective being the improvement of the determination of the ephemerides of Jupiter and the Galilean satellites, by means of providing high-accuracy tracking data of the JUICE spacecraft from which its flight dynamics can be reconstructed. To achieve this, PRIDE will use observations of the spacecraft itself, along with compact extragalactic radio sources by means of near-field phase-referencing very long baseline interferometry (VLBI) observations \citep{2012A&A...541A..43D} performed with networks of ground-based radio telescopes, and Doppler measurements of the spacecraft carrier signal to obtain estimates of the radial velocity of JUICE \citep{2021PASA...38...65M}. The aspects of PRIDE as an operational component of the JUICE mission are presented by \citet{2025CEAS...17..339P}.

During the eight-year-long interplanetary cruise phase, the PRIDE team takes the opportunity to perform occasional VLBI observations of the JUICE spacecraft. Apart from serving as end-to-end practice runs for planning, scheduling, data acquisition, and analysis during the upcoming operations in the Jovian system, these observations could also have scientific merit on their own. For instance, the Venus flyby in August 2025 offered an opportunity for independently validating Venus’ ephemeris solutions \citep{2025CEAS...17..339P}. VLBI data acquired during the flyby have possible impact on two distinct scientific applications. One of them is improved navigation and tests of general relativity \citep{2015CeMDA.123..325F}. Another one is a full test in preparation for JUICE operations in the Jovian system, where the VLBI data will contribute to improving Galilean moon ephemerides during the many flybys around these celestial objects \citep{2024Icar..41616101F}.

The outcome and quality of PRIDE science products rely on VLBI observations of nearby phase-calibrator sources with positions accurately linked to the International Celestial Reference Frame \citep[ICRF,][]{2020A&A...644A.159C}. According to numerical simulations \citep{2025CEAS...17..339P}, the ability of the PRIDE test experiment to improve Venus’ ephemeris accuracy depends critically on the number and quality of phase-reference radio sources available along the track of the spacecraft. Thus, in-beam phase-calibrator sources with positions tied to the ICRF hold a major importance, rendering traditional nodding-style phase referencing observations \citep[e.g.][]{1995ASPC...82..327B} unnecessary, and thus significantly decreasing the idle time spent on antenna slewing. Moreover, the availability of multiple phase-reference sources within a reasonable angular distance ($\sim1\degr$) would contribute to a more accurate astrometry of JUICE.

On average, there is approximately one known extragalactic VLBI source within every $2$ square degrees in the sky \citep[see][]{2025ApJS..276...38P}. However, for the most accurate VLBI position determination of JUICE using in-beam phase referencing \citep[e.g.][]{1999AJ....117.3025F}, arcminute-separation phase calibrators would be required. In such a configuration, both the spacecraft and a strong calibrator source lie within the primary beam of the radio telescopes comprising the VLBI network. Thus, they can be
observed simultaneously, without the need for switching the antennas between the target and the calibrator.

Since the positional accuracy of VLBI phase referencing depends critically on the target--calibrator angular separation \citep[e.g.][]{2020A&ARv..28....6R}, densifying the celestial reference frame around the Ecliptic is of utmost importance for interplanetary spacecraft navigation in general \citep[e.g.][]{2022Univ....8..374D}, and for JUICE in particular, prior to its arrival to the Jovian system. The half-power beamwidth (HPBW) of the primary beam of a $32$-m diameter radio telescope is $4\arcmin$ at $X$ band (around $8.4$~GHz, the JUICE primary downlink frequency). 

Suitable calibrator sources within the HPBW centred on the JUICE trajectory would have allowed in-beam phase-referencing observations with a network of $32$-m or smaller antennas at around the Venus flyby of JUICE. During the 5-day period centred on the date of closest approach to Venus, 31 August 2025, only four known ICRF sources  at $X$ band are found within $2\degr$ of JUICE's trajectory. Additionally, further nine known VLBI-detected sources are listed in the Radio Fundamental Catalog \citep[RFC,][]{2025ApJS..276...38P} within $\sim2\degr$, out of which seven are, in principle, suitable calibrators, i.e. bright and compact enough for phase-referencing observations. To use some of the latter as reference objects for JUICE VLBI measurements, their coordinates should be linked to ICRF by means of phase-referenced relative astrometric observations to nearby ICRF sources. 

With the aim of linking the selected RFC sources to the ICRF in the celestial field that is visible from the ground at around the time of the JUICE Venus gravity assist manoeuvre, and finding possible in-beam calibrator sources within $4\arcmin$ angular distance from the JUICE's trajectory, we conducted nodding-style phase-referencing observations \citep{1995ASPC...82..327B} with the European VLBI Network (EVN). 
In Section~\ref{sec:obs}, we describe the target selection process and the details of the EVN observations. We introduce the methods and steps of data reduction in Section~\ref{sec:data}. Results of the project are presented in Section~\ref{sec:results}. Additionally, a detailed analysis on two interesting target sources is provided in Section~\ref{sec:discussion}. Finally, in Section~\ref{sec:summary}, we give a summary and concluding remarks.

Throughout the paper, a standard flat $\Lambda$CDM cosmological model was used, with parameters $H_0 = 70$~km~s$^{-1}$~Mpc$^{-1}$, $\Omega_\mathrm{M}=0.3$, and $\Omega_\Lambda=0.7$, to calculate projected linear sizes, luminosity distances, and radio powers of extragalactic objects.

\section{Target selection and observations}\label{sec:obs}

Details of the experiment planning and target selection are given in \citet{2025CEAS...17..339P}, here we give a concise summary. To find new potential in-beam calibrator sources, we selected nine radio sources from the Very Large Array Sky Survey \citep[VLASS,][]{2020RNAAS...4..175G,2021ApJS..255...30G} and the mJy Imaging Very Long Baseline Array (VLBA) Exploration \citep[mJIVE,][]{2014AJ....147...14D} surveys within $4\arcmin$ distance along the trajectory of the spacecraft during its Venus flyby (Table~\ref{tab:iccFVC}). While the VLASS sources have never been imaged with VLBI at milliarcsecond (mas) resolution before, the objects taken from the mJIVE list had prior VLBI detection but at about six times lower frequency than at $X$ band, and poorer angular resolution. With the aim to test the detection success rate of sources with unknown total or correlated flux densities at JUICE's downlink frequency, all radio sources along the spacecraft trajectory were considered, without a lower limit on their estimated $8.4$-GHz flux densities.
For the target sources, phase-reference calibrators were predominantly selected from the ICRF. However, in some cases, only calibrators from the RFC were within a reasonable angular distance. Additional observations were carried out to check RFC sources and link them to the ICRF. Thus, three RFC targets within $1\degr$ from the JUICE trajectory were also observed in phase-reference mode (Table~\ref{tab:obs}). For each target, both the in-beam calibrator candidate (ICC) and the RFC sources, two independent phase-reference calibrators (Cal1, Cal2, see Table~\ref{tab:obs}) were selected, to assess the reliability of the relative position determinations.

\begin{table}[h!]
    \centering\small
    \caption{Potential in-beam calibrator sources selected for observations in this work.}
    \label{tab:iccFVC}\setlength{\tabcolsep}{3pt}
    \begin{tabular}{lrcrcc}
     \ttrule
Name	&	$S_\mathrm{F}$~~~~~	&	$S_\mathrm{mJIVE}$	&	$S_\mathrm{V}$	~~~~&	$C_\mathrm{F}$	&	$C_\mathrm{V}$	\\
   & (mJy)~~~~& (mJy) & (mJy)~~~& (beam$^{-1}$) & (beam$^{-1}$) \\
\midrule
J0822+1930	&	17.1$\pm$1.5	&		&	8.2$\pm$0.3	&	0.9	&	0.6	\\
J0828+1908	&	23.6$\pm$1.8	&		&	10.7$\pm$0.3	&	0.9	&	0.7	\\
J0835+1844A	&	41.6$\pm$2.5	&		&	22.5$\pm$0.3	&	0.9	&	0.7	\\
J0835+1844B	&	10.0$\pm$1.2	&		&	6.4$\pm$0.3	&	1.0	&	0.8	\\
J0841+1821	&	7.1$\pm$1.7	&	4.1$\pm$0.6	&	5.0$\pm$0.3	&	1.0 &	1.1	\\
J0841+1826	&	4.3$\pm$1.2	&	1.8$\pm$0.3	&	4.8$\pm$0.3	&	0.9	&	1.0	\\
J0843+1814	&	534.1$\pm$7.6	&	1.2$\pm$0.2	&	241.2$\pm$0.8	&	0.6	&	0.4	\\
J0843+1815	&	3.4$\pm$1.1	&	2.3$\pm$0.4	&	2.1$\pm$0.2	&	1.1	&	1.1	\\
J0843+1816	&	1.5$\pm$0.8	&	1.9$\pm$0.3	&	1.3$\pm$0.3	&	1.1	&	0.8	\\
 \bottomrule
    \end{tabular}
\begin{tabnote}
\textit{Notes:} Col.~1 -- short name, Col.~2 -- 1.4-GHz flux density in FIRST, Col.~3 -- 1.4-GHz flux density in mJIVE, Col.~4 --  3-GHz flux density in VLASS (epoch 2), Col.~5--6 --  kpc-scale compactness (peak intensity to flux density ratio) for FIRST and VLASS, respectively
\end{tabnote}
\end{table}

The VLBI observations were carried out on 23--24 February 2024 with the EVN (project code: EP129, PI: K. Perger) at a central frequency of $8.4$~GHz. The observations lasted for a total of $10$~h with a net on-source time of $7.23$~h. Throughout the project, $8-12$ antennas participated, with Westerbork (WB, The Netherlands), Effelsberg (EF, Germany), Medicina (MC, Italy), Noto (NT, Italy), Onsala (O6, Sweden), Toru\'{n} (TR, Poland), Yebes (YS, Spain), Irbene (IR, Latvia) participating in the observations of all sources (labeled as `base' antennas in Table~\ref{tab:obs}). Depending on their availability, Tianma (T6, China), Nanshan (UR, China), Hartebeesthoek (HH, South Africa), and Kunming (KM, China) were also included in the most complete setups. The data were recorded in left and right circular polarizations at a total rate of $2048$~Mbit~s$^{-1}$. A total of eight intermediate frequency channels (IFs) were used, with $64$ spectral channels in each. The total bandwidth per polarisation per IF was $32$~MHz. The data were correlated at the EVN Data Processor at the Joint Institute for VLBI European Research Infrastructure Consortium (JIVE, Dwingeloo, the Netherlands) with a $2$~s averaging time. The names of calibrator sources, details on the participating antennas, and net on-source times are listed in Table~\ref{tab:obs}.

To investigate the kpc-scale radio structure and the environment of J0843+1814,  archival Very Large Array (VLA) A-configuration data were used, from observations at $1.5$~GHz (L-band, AV162, PI: W. van Breugel, 9 December 1988) and $4.9$~GHz (C-band, LAWR, PI: C.~R. Lawrence, 11--12 February 1982). These data were obtained from the  US National Radio Astronomy Observatory (NRAO) archive\footnote{\url{https://data.nrao.edu/}}. 
In the L-band observations, $26$ antennae participated, with a total observing time of $12$~h~$57$~min, and a net on-target time of $8.5$~min. Two IFs were used with a bandwidth of $25$~MHz. The absolute flux density calibrator was 3C~286, while gains were calibrated using the sources B0851+202 (J0854+2006) and B0741$-$063 (J0744$-$0629). The C-band observations lasted for a total of $24$~h~$24$~min, with $22$ antennae participating. The single IF had $50$~MHz bandwidth. The target was observed with net on-source time of $3$~min, with 3C~286 and B0839+187 (J0842+1835) as the flux density and gain calibrators, respectively. In both experiments, the data were recorded in left and right circular polarisations.

\section{Data reduction}\label{sec:data}
The EVN data were calibrated in the NRAO Astronomical Image Processing System package \citep[\textsc{aips},][]{2003ASSL..285..109G} in a standard way \citep[e.g.][]{1995ASPC...82..227D}. After starting with an a priori amplitude calibration based on the antenna system temperatures and gain curves, ionospheric delay and parallactic angle phase corrections were carried out. Then a manual phase calibration was performed on a $1$-min data segment of a bright fringe-finder source (J0830+2410), followed by data inspection and flagging. For each calibrator source, separate global fringe-fitting runs \citep{1983AJ.....88..688S} were performed. Data for each ICC source were phase-referenced in two separate ways to their respective calibrators, while RFC sources were calibrated three times, including the two phase-referencing runs (Cal1,Cal2), and fringe fitting based on their own data as well (SF). 

The calibrator and target source visibility data were then exported into the \textsc{Difmap} program \citep{1997ASPC..125...77S} for imaging and model fitting. ICC sources were imaged using the \textsc{clean} deconvolution method \citep{1974A&AS...15..417H}. For RFC and ICRF sources, hybrid mapping was carried out, with alternate cycles of \textsc{clean} image deconvolution and phase and amplitude self-calibration. The final \textsc{clean} images were then imported to \textsc{aips}, and coordinates of the brightness peaks were determined using the \textsc{maxfit} function. Flux densities of the sources were determined by fitting Gaussian model components directly to the visibility data in \textsc{Difmap}. Alternatively, to assess coherence loss during phase-reference imaging, circular Gaussian brightness distribution model fitting was performed directly to the \textsc{aips}-calibrated visibility data sets of RFC sources (Cal1, Cal2, and SF) without self-calibration.

Astrometric uncertainties were estimated considering the following factors:
\begin{itemize}
 \item Position error of the respective calibrator source(s), obtained from the ICRF or RFC catalogues.
 \item Errors arising from the signal-to-noise ratio (SNR) of the target source \citep{1990AJ.....99.1663L}.
 \item Residual systematic errors, depending on the deconvolved source size, number of telescopes, and observing time \citep{2001ApJ...550..287C}.
 \item Uncertainty arising from the target--calibrator separation \citep{2006A&A...452.1099P}.
 \item Target position error \citep{2008AJ....136..159L}.
\end{itemize}
The final astrometric uncertainties were calculated by adding the individual error components in quadrature, and are given in parentheses in Columns~3--4 in Table~\ref{tab:model_target} and Columns~2--3 in Table~\ref{tab:model_rfc}, for ICC and RFC sources, respectively. Image and model parameter uncertainties were calculated following \citet{2008AJ....136..159L}.

The archival VLA data for J0843+1814 were calibrated in \textsc{aips} in a standard way, while imaging and model fitting were performed in \textsc{Difmap}.

\section{Results}\label{sec:results}

\subsection{ICC sources detected}
Out of the nine targeted ICC sources, four were detected in the phase-referencing observations, resulting in a detection rate of $44\%$. Typical VLBI detection rates in wide-field surveys are $\sim20-30\%$ \citep[e.g.][]{2013A&A...550A..68C,2013A&A...551A..97M,2014AJ....147...14D,2017A&A...607A.132H,2018A&A...619A..48R,2023MNRAS.519.1732N}. The somewhat higher success of detecting the targets in our project can be attributed to several factors. On the one hand, this is a small sample drawn from a small region, and the sources are not distributed homogeneously over the sky, so there must be a random factor. On the other hand, our higher detection rate can result from better pointing, since more than half of the sample had prior VLBI detection in mJIVE \citep{2014AJ....147...14D}. As all the ICC sources were expected to have mJy-level flux densities at mas scale, the true positions of those without mJIVE detection could have been far off the phase centres (up to $\sim500$~mas, the astrometric precision of VLASS). Thus, their faint emission could have been missed due to the higher noise levels. Indeed, our detected mJIVE-selected sources are all found within a couple mas from the targeted a priori positions. 

\begin{figure}[htp]
 \centering
 \includegraphics[width=0.9\linewidth]{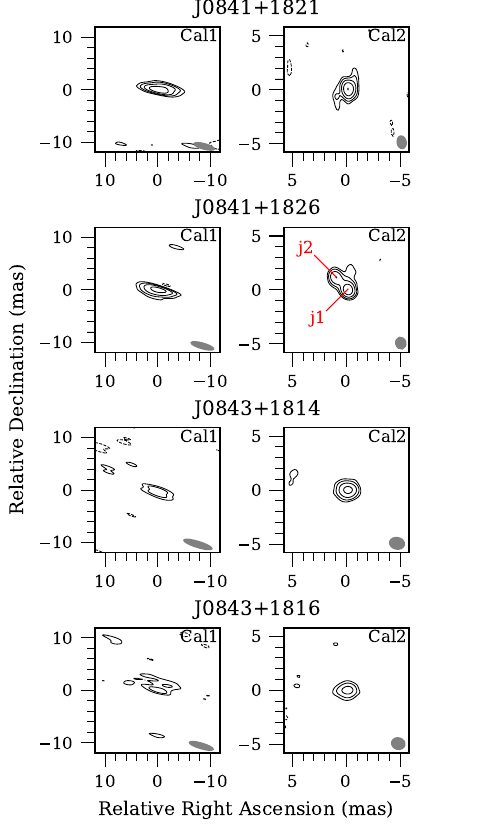}
 \caption{EVN radio maps at $8.4$~GHz of the detected ICC sources. 
 The phase reference-calibrator used to perform the fringe fitting is indicated in the upper right corner for each map. These calibrators are listed in Table~\ref{tab:obs}. The first contours are drawn at $\pm2.5\sigma$ off-source rms noise, and the positive levels increase by a factor of 2. The restoring beams are denoted in the bottom right corners. Image parameters are listed in Table~\ref{tab:image}.}
 \label{fig:target}
\end{figure}

The \textsc{clean} maps of the detected ICC sources are shown in Figure~\ref{fig:target}, separately for the respective calibrator sources Cal1 and Cal2, as detailed in Table~\ref{tab:obs} and denoted in the top right corners. Image parameters are listed in Table~\ref{tab:image}. Source coordinates and model parameters (flux density and Gaussian model component size, full width at half maximum, FWHM) are listed in Table~\ref{tab:model_target}.

The detected ICC sources typically show compact, single-component features, except for J0841+1826, which has a core--jet-like structure extended towards the northeastern direction. It is recovered by the observations with Cal2, thanks to the long baselines and the nearly circular restoring beam in the experiment. The sum of the flux densities of the two components is $S_\mathrm{2,total}=2.0 \pm 0.3$~mJy, which agrees with the single-component flux density from the Cal1 experiment, $S_1=1.8\pm0.4$~mJy, where the source remains unresolved with the larger restoring beam.

All four ICC sources detected in the observations were previously detected with VLBI in the $1.4$-GHz mJIVE survey \citep{2014AJ....147...14D}. The power-law spectral indices ($\alpha$) between the frequencies ($\nu$) of $1.4$ and $8.4$-GHz, based on the mJIVE and our EVN flux densities ($S$), were calculated following the $S \sim \nu^\alpha$ convention, and are listed in Table~\ref{tab:model_target}. The spectral index for the Cal2 observations of J0841+1826 was calculated using the total flux density value. The spectra of the compact VLBI-detected features are generally flat ($\alpha\le-0.5$) for all detected sources, except for J0843+1816. However, considering the uncertainties, this source can also be regarded as having a flat spectrum. 

Although the compactness of the detected ICC targets would make them suitable as a calibrator, their low (mJy- and sub-mJy-level) flux densities (Table~\ref{tab:model_target}) prevent them from being used as phase-reference sources for spacecraft navigation in practice. While they are not bright enough for producing sufficiently high SNR ($\mathbf{\gtrsim10}$) for fringe fitting within the atmospheric coherence time ($\sim2$~min  at $8.4$~GHz), since their coordinates are now linked to ICRF, they could possibly be used as quality check sources if observed within the antenna primary beams together with the spacecraft in a traditional nodding-style phase-referencing experiment. In case the coordinates determined for the in-beam check source in such project would differ significantly from their known values (i.e. from mJIVE or this work; accurate to $\lesssim 0.3$~mas, Table~\ref{tab:model_target}), such a difference would likely result from the phase-referencing procedure, and could be accounted for prior to being used to improve the spacecraft position solution.

\subsection{ICC sources not detected}

The remaining five ICC targets (J0822+1930, J0828+1908, J0835+1844A, J0835+1844B, and J0843+1815) were not detected with the EVN at $8.4$~GHz with any of their two phase-reference calibrators (Table~\ref{tab:obs}). Upper limits of their flux densities are given in Table~\ref{tab:model_target} based on the $5\sigma$ image noise, assuming a compact source unresolved with the interferometer. Only one of them, J0843+1815, had prior VLBI detection at $1.4$~GHz in the mJIVE survey \citep{2014AJ....147...14D} where its image revealed an extended structure.

There appears no clear connection between the complexity of the kpc-scale radio structure and the success of a $8.4$-GHz VLBI detection in our sample of ICC targets. Both the detected and non-detected sources show a variety of morphologies at either or both of the frequencies of $1.4$ and $3$~GHz, in the images of the Faint Images of the Radio Sky at Twenty-centimeters \citep[FIRST,][]{1997ApJ...475..479W,2015ApJ...801...26H} survey and VLASS \citep{2020RNAAS...4..175G,2021ApJS..255...30G}, respectively. In general, we found that the studied ICC sources present themselves in FIRST and VLASS on arcsec scales as single point-like components, or slightly extended sources, or those showing  double or multi-component structure. For comparison, their flux densities and compactness parameters are collected in Table~\ref{tab:iccFVC}. Similarly, no connection was found between the pc-scale detection and the kpc-scale radio spectral index. The majority of the sources, both VLBI-detected and non-detected, have steep spectra at kpc scales. Moreover, the success of our VLBI detection does not seem to have any dependence on the kpc-scale compactness, i.e. the ratio of the image peak intensity and the flux density ($C\equiv P_\mathrm{kpc}/S_\mathrm{kpc}$). Sources with rather extended structure and compactness as low as $C=0.4$~beam$^{-1}$ were recovered with the EVN, whilst others with $C=(0.8-1.1)$~beam$^{-1}$ remained undetected or were resolved out on mas scale. As the majority of the successfully detected sources are in the order of magnitude of $S \sim 1$~mJy, it is possible that longer on-source integrations, thus better thermal sensitivity would recover fainter compact pc-scale emission for the rest of the ICC targets.

\subsection{RFC sources}
\begin{figure}[ht!]
 \centering
 \includegraphics[width=0.98\linewidth]{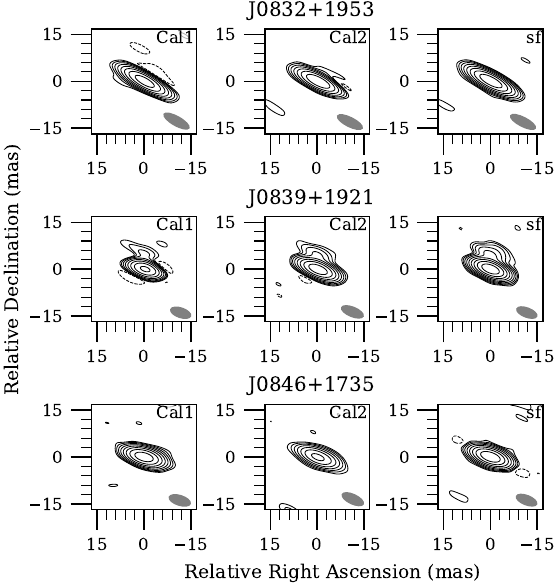}
 \caption{EVN radio maps at $8.4$~GHz of the detected RFC sources. The left and middle panels show data acquired using different phase-reference calibrators to perform the fringe-fitting. The names of the respective calibrators are listed in Table~\ref{tab:obs}. The right panel shows the image acquired after fringe-fitting to the data of the targeted RFC source itself (sf -- self-fringe). The first contours are drawn at $\pm3\sigma$ off-source rms noise, and the positive levels increase by a factor of 2. The restoring beams are denoted in the bottom right corners. Image parameters are listed in Table~\ref{tab:image}.}
 \label{fig:rfc}
\end{figure}

All three of the RFC sources were detected in both of their phase-referencing experiments, as well as in their self-fringe-fitted data sets. The \textsc{clean} maps obtained from the hybrid mapping are shown in Figure~\ref{fig:rfc}, corresponding to data calibrated with the respective phase-reference sources from Table~\ref{tab:obs}. Image properties of the \textsc{clean} maps are listed in Table~\ref{tab:image}.
Source coordinates and parameters of the fitted Gaussian models are given in Table~\ref{tab:model_rfc}.

The estimated errors of the modeled flux densities of the RFC sources in our EVN measurements are $\sim15\%$ (Table~\ref{tab:model_rfc}), including a general $10\%$ VLBI absolute amplitude calibration uncertainty added to the thermal noise contribution in quadrature. As no formal uncertainties are given for the median flux density values in the RFC catalogue \citep{2025ApJS..276...38P}, we considered relative uncertainties of the same order of magnitude as for the EVN observations. With this, our model-fitted flux densities agree with the short-baseline RFC catalogue values within the uncertainties.

\subsection{Positional uncertainties and coherence loss}\label{sec:errors}

\begin{figure}[ht]

 \centering
 \includegraphics[width=0.88\linewidth]{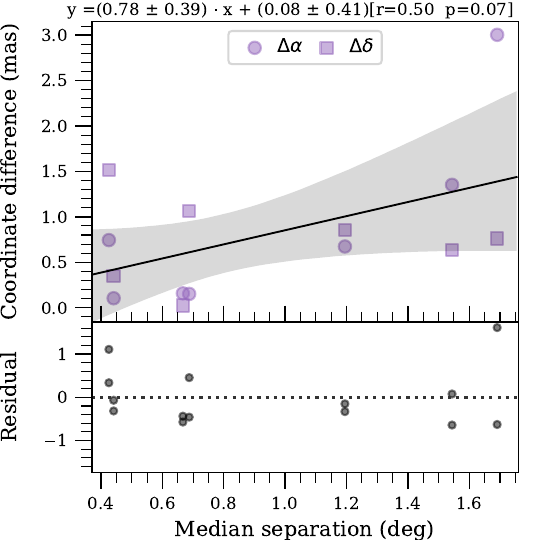}
 \caption{Positional difference in the right ascension and declination of the target ICC and RFC sources, determined from the \textsc{clean} maps resulting from the Cal1 and Cal2 data sets. The black line and the grey area denote the fitted linear equation and the $95\%$ confidence interval, respectively, resulting from the Pearson correlation test. Parameters of the linear fit are given on top of the figure, while residuals are shown in the lower panel.}
 \label{fig:diff}
\end{figure}

For the majority of the targets, the difference between the coordinates obtained with the Cal1 and Cal2 phase-reference calibrators is generally larger in declination than in right ascension (see Figure~\ref{fig:diff}). This could arise from the fact that all detected ICC and RFC targets were observed in parts of the experiment where the phase referencing was conducted with the calibrators J0839+1802 and J0842+1835. Both of them have a complex structure, extended primarily in the north--south direction (see Figure~\ref{tab:cal} in Appendix\,\ref{sec:calibrators}). This can be a contributing factor for the positional differences exceeding the estimated formal uncertainty values. On the other hand, no correlation was found between the right ascension or declination differences and the target--calibrator separations. The Pearson linear correlation test resulted in a coefficient of $r=0.5$ and a $p$ value of $0.07$ (Figure~\ref{fig:diff}), suggesting a weak positive correlation, however, due to the marginally high $p$ value and the limited size of the sample, it cannot be considered statistically significant.  On the other hand, the coordinate differences may be affected by other factors than just the target--calibrator separation (Figure~\ref{fig:diff}). Note that, at least for the ICC targets, the Cal1 and Cal2 solutions come from EVN subarrays that are significantly different in extent (see Table~\ref{tab:obs}).  Obviously, these provide different angular resolutions and position determination accuracies (see Table~\ref{tab:image} and the restoring beams in Figure~\ref{fig:target}). Nevertheless, the positions determined from data using Cal1 and Cal2 solutions are still consistent with each other within $\mathbf{2\sigma}$ level of the formal errors, implying that the uncertainties estimated from theoretical considerations might somewhat underestimate the practical values.

\begin{figure}[ht]
\centering
\includegraphics[width=0.88\linewidth]{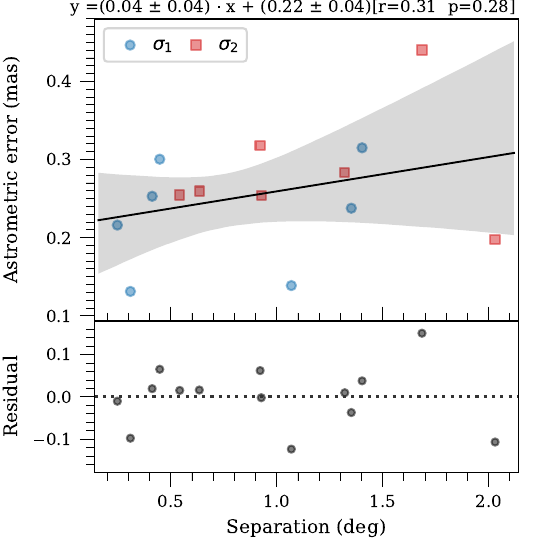}
\caption{Astrometric uncertainties of the observations as a function of target--calibrator separation. Blue circles and red squares show values calculated for data sets calibrated with Cal1 and Cal2, respectively. The solid black line and grey area denote the fitted linear regression equation and a $95\%$ confidence interval, respectively, for all data points (blue+red). Parameters of the linear fit are given on top of the figure, while residuals are shown in the lower panel. } \label{fig:astrometry}
\end{figure}

We found that quadratic errors (arising from astrometric uncertainties, as defined in Section~\ref{sec:data}) do not show significant correlation with the angular distance between the target and calibrator sources (with $r=0.37$ and $p=0.22$; see Figure~\ref{fig:astrometry}), contrary to the findings of e.g. \citet{2004ApJ...604..339C,2017AJ....153..105R}, who showed an approximately linear relation between these parameters, up to $\sim6\degr$ separations. This apparent contradiction can be easily resolved by the facts that the largest target--calibrator separation in our experiment was only $2.26\degr$, and that our observations were carried out at a higher frequency where the dependence of the astrometric error on the target--calibrator separation is expected to be weaker \citep[see e.g.][]{2001ApJ...550..287C}. We also found that neither the individual error components, nor the individual properties influencing the error components (e.g. SNR, Gaussian model FWHM) correlate with the angular separation from the respective calibrator sources. 

Although the contributions from individual astrometric error components are slightly different from target to target, there is a clear trend in their relative weights in the total error budget. In this sample, the positional uncertainties of the calibrators and targets, and the systematic errors dominate the error budget, each of the three being responsible for $\sim30\%$ of the total value. The uncertainties arising from the target--calibrator separation, and the target SNR components collectively contribute only $\lesssim10\%$ of the total budget, with the former exerting the least influence, being responsible for $\lesssim2.5\%$ or the errors.

Source flux densities measured in phase-referenced VLBI experiments are reduced by signal decoherence \citep{2010A&A...515A..53M}. In our experiment, in the case of the RFC targets, we were able to estimate the amount of this coherence loss in practice, since both phase-referenced and self-fringe-fitted observations could be analysed. By comparing the model-fitted flux densities for non-self-calibrated data, we found that the average reduction caused by coherence loss was $\sim20\%$. This is broadly consistent with values reported for earlier EVN experiments at different frequencies \citep[e.g.][]{2006A&A...445..413M,2019A&A...630L...5G,2024A&A...690A.321K}. Therefore, in estimating the true flux densities and detection upper limits for the weak ICC sources (Table~\ref{tab:model_target}), this coherence loss should ideally be corrected for.

\begin{table*}[ht]
 \centering\small\caption{Details on the observations.}\label{tab:obs}
 \begin{tabular}{llllllllll}
\ttrule																	
Name	&	Origin	&	Calibrator 1	&	Calibrator 2	&	$d_1$&	$d_2$&	Antennae 1 	&	Antennae 2	&	$t_1$ 	&	$t_2$\\

&&&& ($\degr$) & ($\degr$) &&& (s) & (s)\\\midrule
\multicolumn{9}{c}{ICC sources}\\\midrule							J0822+1930	&	V	&	J0817+1958	&	J0829+1754	&	1.49	&	2.26	&	base+HH,UR,KM,T6	&	base+HH,UR,KM,T6	&	707	&	707	\\
J0828+1908	&	V	&	J0832+1832	&	J0829+1754	&	1.09	&	1.24	&	base+HH,UR,KM,T6	&	base+HH,UR,KM,T6	&	708	&	825	\\
J0835+1844A	&	V	&	J0832+1832	&	J0839+1802	&	0.78	&	1.27	&	base+HH,UR,KM,T6	&	base+HH,UR,KM,T6	&	708	&	708	\\
J0835+1844B	&	V	&	J0832+1832	&	J0839+1802	&	0.92	&	1.14	&	base+HH,UR,KM,T6	&	base+HH,UR,KM,T6	&	707	&	707	\\
J0841+1821	&	V/m	&	J0842+1835	&	J0839+1802	&	0.25	&	0.63	&	base+HH	&	base+HH,UR,KM	&	707	&	707	\\
J0841+1826	&	m	&	J0842+1835	&	J0839+1802	&	0.31	&	0.54	&	base+HH	&	base+HH,UR,KM	&	590	&	707	\\				
J0843+1814	&	m	&	J0842+1835	&	J0839+1802	&	0.45	&	0.93	&	base+HH	&	base+HH,UR	&	590	&	708	\\
J0843+1815	&	m	&	J0842+1835	&	J0839+1802	&	0.50	&	1.03	&	base+HH	&	base+HH,UR	&	590	&	708	\\
J0843+1816	&	m	&	J0842+1835	&	J0839+1802	&	0.41	&	0.92	&	base+HH	&	base+HH,UR	&	826	&	825	\\
\midrule\multicolumn{9}{c}{RFC sources}\\\midrule
J0832+1953	&	RFC	&	J0832+1832	&	J0836+2139	&	2.03	&	1.35	&	base	&	base	&	115	&	174	\\
J0839+1921	&	RFC	&	J0842+1835	&	J0839+1802	&	1.07	&	1.32	&	base	&	base	&	116	&	115	\\
J0846+1735	&	RFC	&	J0842+1835	&	J0839+1802	&	1.40	&	1.69	&	base	&	base	&	58	&	116	\\
\bottomrule														 \end{tabular}\begin{tabnote}
{\textit{Notes:} Col.~1 -- short name of the target, Col.~2 -- original catalogue for source selection (V = VLASS, m = mJIVE, RFC), Col.~3--4 -- first (Cal1) and second (Cal2) phase-reference calibrator, Col.~5--6 -- distances between the target and its first and second phase-reference calibrators, Col.~7--8 -- antennae observing in the first and second rounds of scans; `base' denotes stations which participated in all observations: WB, EF, MC, NT, O6, TR, YS, and IR,  Col.~9--10 -- total target on-source times in the first and second rounds of scans.\vspace{10pt}}
\end{tabnote}
\end{table*}

\begin{table*}[ht]\centering\small
\caption{Image parameters for Figures~\ref{fig:target}--\ref{fig:rfc}.}\label{tab:image}
\setlength{\tabcolsep}{1.5pt}
 \begin{tabular}{cccccccccccccccc}
 \ttrule																														
Name	&	$P_1$	&	rms$_1$	&	$b_\mathrm{min1}$	&	$b_\mathrm{maj1}$	&	$b_\mathrm{PA1}$	&	$P_2$	&	rms$_2$	&	$b_\mathrm{min2}$	&	$b_\mathrm{maj2}$	&	$b_\mathrm{PA2}$	&	$P_\mathrm{SF}$	&	rms$_\mathrm{SF}$	&	$b_\mathrm{min,SF}$	&	$b_\mathrm{maj,SF}$	&	$b_\mathrm{PA,SF}$	\\

&(mJy & (mJy & (mas) & (mas) & ($\degr$) & (mJy & (mJy & (mas) & (mas) & ($\degr$) & (mJy & (mJy & (mas) & (mas) & ($\degr$)\\
&beam$^{-1}$) & beam$^{-1}$) &  &  & & beam$^{-1}$) & beam$^{-1}$) &  &  & & beam$^{-1}$) & beam$^{-1}$) &  &  &\\

\midrule\multicolumn{16}{c}{ICC sources}\\\midrule						J0841+1821	&	$1.0\pm0.2$	&	$0.03$	&	1.0	&	4.0	&	75	&	$1.1\pm0.2$	&	$0.03$	&	0.8	&	1.2	&	9	& ... & ... & ... & ... & ... \\				
J0841+1826	&	$1.1\pm0.2$	&	$0.04$	&	1.0	&	4.4	&	74	&	$1.1\pm0.2$	&	$0.04$	&	0.9	&	1.1	&	15	& ... & ... & ... & ... & ... \\				
J0843+1814	&	$0.6\pm0.2$	&	$0.07$	&	1.1	&	5.6	&	73	&	$0.6\pm0.1$	&	$0.03$	&	1.1	&	1.4	&	78	& ... & ... & ... & ... & ... \\			
J0843+1816	&	$0.3\pm0.1$	&	$0.03$	&	1.0	&	4.7	&	74	&	$0.3\pm0.1$	&	$0.02$	&	1.1	&	1.2	&	76	& ... & ... & ... & ... & ... \\									
\midrule\multicolumn{16}{c}{RFC sources}\\\midrule												
J0832+1953	&	$45.8\pm1.9$	&	$0.08$	&	2.8	&	9.0	&	61	&	$46.8\pm2.5$	&	$0.13$	&	2.9	&	8.7	&	64	&	$51.3\pm2.0$	&	$0.08$	&	2.8	&	8.8	&	63	\\

J0839+1921	&	$31.9\pm1.6$	&	$0.08$	&	3.0	&	6.6	&	71	&	$34.7\pm1.4$	&	$0.06$	&	2.9	&	7.3	&	68	&	$43.1\pm1.2$	&	$0.03$	&	2.9	&	6.7	&	69	\\
J0846+1735	&	$30.5\pm1.2$	&	$0.05$	&	3.1	&	7.0	&	72	&	$29.3\pm1.9$	&	$0.12$	&	3.0	&	7.3	&	65	&	$42.5\pm1.6$	&	$0.06$	&	2.9	&	6.9	&	69	\\

\bottomrule
 \end{tabular}
\begin{tabnote}
{\textit{Notes:} Col.~1 -- short name, Col.~2 -- peak intensity (Cal1), Col.~3 -- rms noise (Cal1), Col.~4--6 -- minor and major axis FWHM size and major axis position angle of the restoring beam (Cal1), Col.~7 -- peak intensity (Cal2), Col.~8 -- rms noise (Cal2), Col.~9--11 -- minor and major axis FWHM size and major axis position angle of the restoring beam (Cal2), Col.~12 -- peak intensity (SF), Col.~13 -- rms noise (SF), Col.~14--16 -- minor and major axis FWHM size and position angle of the restoring beam (SF). Cal1 and Cal2 denote data sets fringe-fitted with calibrators listed in Table~\ref{tab:obs}, while SF refers to the self-fringe-fitted data for the bright RFC sources only.}
 \end{tabnote}
\end{table*}

\section{Discussion of individually interesting sources}\label{sec:discussion}
\subsection{J0841+1826 -- the only ICC target with resolved mas-scale structure}
Based on its arcsec-scale FIRST and VLASS imaging, this radio source has an unresolved structure confined within $\sim1$~kpc projected linear size. Its spectroscopic redshift, $z_\mathrm{sp}=0.4622\pm0.0001$ \citep[][]{2020ApJS..249....3A}, is taken from the 16th Data Release (DR16) of the Sloan Digital Sky Survey (SDSS). The flux densities of J0841+1826 measured at the three different VLASS epochs agree within errors, with $S_\mathrm{median} = (4.5\pm0.3)$~mJy.

Among our two phase-referenced EVN observations, one has much better resolution (Cal2, Figure~\ref{fig:target}). There the mas-scale source  structure was resolved into two components (j1 and j2) which can be interpreted as either a core--jet or a double-jet morphology. Notably, the flux densities of the two components agree with each other within errors (Table~\ref{tab:model_target}). Below we speculate about the possible interpretations of this structure.

The apparent brightness temperature was calculated for both components by applying the formula \citep{1982ApJ...252..102C}
\begin{equation}
    T_{\rm{b}}=1.22\times {10}^{12}\,(1+z)\,\frac{S}{{\vartheta }^{2}{\nu }^{2}}\,{\rm{K},}
\end{equation}
where $S$ is the flux density in Jy, $\vartheta$ the angular size of the fitted circular Gaussian model component (FWHM) in mas, and $\nu$ the observing frequency in GHz. In the Cal2 experiment, where the structure was resolved into two components, the upper limit on the modeled size of j1 (Table~\ref{tab:model_target}) resulted in a lower limit on the brightness temperature of $T_{\rm{b,j1}}>2.8\times10^8$~K, while for the j2 component, the value of $T_{\rm{b,j2}}=(2.4\pm1.0)\times10^8$~K was found. For the Cal1 experiment, where the individual components were not resolved, a lower limit of  $T_{\rm{b,j1+j2}}>1.2\times10^9$~K was found, while the value estimated from the $1.4$-GHz mJIVE data \citep{2014AJ....147...14D} is similar, with $T_{\rm{b,mJIVE}}>3.8\times10^8$~K. Based on the brightness temperature values, neither component shows relativistic beaming, implying a large jet viewing angle with respect to the observer's line of sight. 

Monochromatic radio powers were estimated using the 

\begin{equation}
P_\nu=4\pi\,D_\mathrm{L}^2\,S_\nu\,(1+z)^{-\alpha-1},
\end{equation}

formula, where $S_\nu$ is the sum of the component flux densities at the given frequency in W~Hz$^{-1}$~m$^{-2}$, $ D_\mathrm{L}$ the luminosity distance at redshift $z$ in meters, and $\alpha$ the spectral index. The spectral index $\alpha=-0.1\pm0.2$ was calculated using the $1.4$-GHz mJIVE and the $8.4$-GHz EVN observations. The $8.4$-GHz radio powers are moderate, being $P_\mathrm{8.4GHz,Cal1}=6.8\cdot10^{20}$~W~Hz$^{-1}$ and $P_\mathrm{8.4GHz,Cal2}=1.1\cdot10^{21}$~W~Hz$^{-1}$ for the unresolved j1 component in Cal1 and for the total system (j1+j2) in Cal2 experiments, respectively. A similar value was found at $1.4$~GHz ($P_\mathrm{mJIVE}=1.2\cdot10^{21}$~W~Hz$^{-1}$). 

The separation of the j1 and j2 components translates to a projected linear size of $8.8$~pc, and neither of them shows Doppler enhancement. These properties favour the double-jet morphology, making J0841+1826 a candidate for a compact symmetric object \citep[CSO, e.g.][]{1994ApJ...432L..87W,2014A&A...569A..22M,2021AN....342.1185R,2024ApJ...961..241K}. CSOs are jetted active galactic nuclei (AGN) that could potentially evolve into large radio galaxies \citep{2012ApJ...760...77A}. The radio power of $\sim10^{21}$~W~Hz$^{-1}$ is similar to those of faintest CSOs \citep{2024ApJ...961..242R}. This possible interpretation is further supported by the estimated value of the jet inclination angle, which was found to be $\sim85\degr$, based on the flux-density ratio -- inclination relation \citep[e.g.][]{1985ApJ...295..358L,2005AIPC..745..510D,2019RMxAA..55..151A}. A precise optical position of J0841+1826 would further support the CSO nature \citep[see][]{2025A&A...704A..93A}, however, the source does not have an optical counterpart detected with the Gaia space astrometry mission  \citep{2016A&A...595A...1G,2023A&A...674A...1G}.

On the other hand, if we consider that the calculated brightness temperature for j1 is a lower limit, the presence of Doppler boosting cannot be formally excluded. Alternatively, j1 and j2 could be interpreted as a core--jet structure in a low-luminosity AGN \citep[LLAGN, e.g.][]{2002A&A...392...53N,2023A&ARv..31....3B,2023A&A...670A..22F}. The low level of the monochromatic radio powers and thus the derivable radio luminosities ($ \nu P \approx 10^{31}$~W), and the extent of the pc-scale structure are similar to those found in nearby LLAGN \citep{2000ApJ...542..197F,2002A&A...392...53N,2014ApJ...787...62M,2025ApJS..277...56C}. Additionally, considering the brighter j1 component as the AGN `core', the spectral index of the source is also comparable to the typical flat spectra of LLAGN \citep{2002A&A...392...53N,2014ApJ...787...62M,2018MNRAS.476.3478B}.
 
\subsection{J0843+1814 -- a young radio galaxy in the centre of a cluster}
\begin{figure}[h!]
    \centering
    \includegraphics[width=\linewidth]{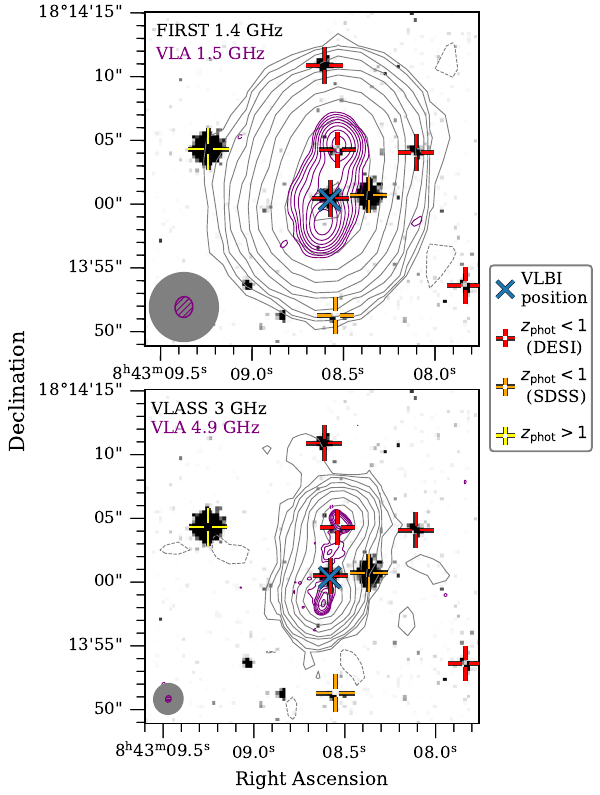}
    \caption{Kpc-scale radio structure of J0843+1814 (contours) and its environment in the optical (grayscale). The restoring beam sizes of the different radio observations are shown in the bottom left corners of each panel. The blue X marks the position of the VLBI-detected compact source, the red and orange crosses denote sources within $15\arcsec$ with photometric redshifts in SDSS and DESI of $z_\mathrm{phot}<1$, while the yellow cross  marks the only $z_\mathrm{phot}>1$ object in the field, respectively. \textit{Top panel:} L-band contour maps overlayed on the DESI $z$-band optical image. Black and purple contours represent the FIRST observations and the VLA-A image (experiment AV162), respectively. \textit{Bottom panel:} Higher-frequency contour maps overlayed on the DESI $z$-band image. Black and purple contours represent the VLASS image mosaic and the VLA-A image (experiment LAWR), respectively.}
    \label{fig:14kpc}
\end{figure}

In sharp contrast to the simple VLBI (pc-scale) morphology of J0843+1814 showing a compact, unresolved emission region, the kpc-scale structure and environment of the source paint a more interesting picture. The nature of the radio source is already hinted by its elongated shape in the $5\arcsec$ angular resolution FIRST radio map, with the $0\farcs5$-resolution VLASS images resolving the structure into a classical double-lobed radio galaxy (Figure~\ref{fig:14kpc}). Indeed, investigating archival data obtained with the VLA in its most extended A configuration (VLA-A), J0843+1814 shows a classical Fanaroff--Riley II \citep[FRII,][]{1974MNRAS.167P..31F} radio galaxy morphology (Figure~\ref{fig:14kpc}).

The positions of the fitted model components of the L- and C-band VLA-A images agree within uncertainties.  We note that the northern region is somewhat an exception, as the sole L-band component was resolved, and three model components were needed to characterise the C-band brightness distribution.

To test the FRII classification, radio powers were estimated, using the photometric redshift of $z_\mathrm{ph}=0.752$ acquired from Dark Energy Spectroscopic Instrument (DESI) Legacy Imaging Survey Photometric Redshift Catalogue \citep[8th data release,][]{2022MNRAS.512.3662D}. The power-law spectral index is $\alpha=-1.1$, following the $S_\nu\sim\nu^\alpha$ convention. The resulting radio powers of $P_\mathrm{1.5GHz}=1.2\times10^{27}$~W~Hz$^{-1}$ and $P_\mathrm{4.9GHz}=3.2\times10^{26}$~W~Hz$^{-1}$ support the FR II classification.

The projected linear (hot spot-to-hot spot) extent of the radio emission spans $\sim50$~kpc. With this size, J0843+1814 belongs to the smaller FRII galaxies \citep[e.g.][]{2011AJ....141...85R,2008MNRAS.390..595M,2017A&A...601A..81C}, and is slightly larger than the radio galaxy precursor FR0 objects \citep[e.g.][]{2019A&A...627A.108J,2023A&ARv..31....3B}. This implies that J0843+1814 might be a young, recently formed radio galaxy.

Within $15\arcsec$ separation from the VLBI core, six optical point sources (Figure~\ref{fig:14kpc}) are identified in the DESI photometric redshift (photo-$z$) catalogue. The closest optical source is located at the VLBI (core) position within uncertainties. Coincidentally, the next closest ($\rho=3\farcs3$, $\sim25$~kpc projected linear distance) is near to the northern hot spot. Due to the large number of close optical sources within a few kpc in projection onto the sky, we investigated their relation and the possibility of clustering. Out of the six optical sources, five have photometric redshifts of $z_\mathrm{phot}<1$, including the target object. The photometric redshifts of the five sources agree within errors, with a median value of $z_\mathrm{phot,med}=0.752\pm0.117$.
We note that out of the four additional sources visible in the DESI image but not included in the photo-$z$ catalogue (one towards the west and three to the south of the VLBI position), two have available counterparts in the SDSS, with photometric redshifts of $z_\mathrm{ph,W}=0.914\pm0.045$ and $z_\mathrm{ph,S}=0.439\pm0.157$ \citep{2015ApJS..219...12A}.

The physical distances of the DESI sources were investigated considering both transverse and radial separations. Based solely on the transverse separations of the presumed members, the distances vary between $28-153$~kpc, with a median value and standard deviation of $76\pm34$~kpc. Considering both the small-scale \citep[$10-200$~kpc~h$^{-1}$,][]{2012MNRAS.424.1363K,2017MNRAS.468...77E,2019MNRAS.485.2026B} and large-scale lengths \citep[$6-12$~Mpc~h$^{-1}$][]{2007ARep...51..820P,2008MNRAS.391.1674W,2011MNRAS.418.2251F,2017MNRAS.464.3271M,2017A&A...600A..97R} found for low-redshift AGN, the five sources might belong to a common cluster. However, by estimating physical distances including radial separations (based on the redshift differences), values between $100-500$~Mpc were found, which would, in principle, disagree with the clustering hypothesis. However, the uncertainties and differences of the photometric redshifts are in the same order of magnitude. On the other hand, large quasar groups with characteristic sizes up to $350$~Mpc~h$^{-1}$ were also identified at $z\lesssim1$ \citep[e.g.][]{2003MNRAS.344..283B,2004A&A...421..157H,2007ARep...51..820P,2012MNRAS.419..556C,2013MNRAS.429.2910C,2016MNRAS.461.2267M}. 

No X-ray point source can be associated with the target within search radii $\sim100\arcmin$, $40\arcmin$, and $4\arcmin$ in the Chandra \citep{2010ApJS..189...37E},  ROSAT \citep{1998AN....319..347F,2000IAUC.7432....3V,2016A&A...588A.103B} and XMM-Newton \citep{2016A&A...590A...1R,2020A&A...641A.137T}, and the eRosita \citep{2024A&A...682A..34M} survey catalogues. On the other hand, an extended X-ray emitting region overlaps with the field of J0843+1814 in the broad-band $1-2.3$~keV eRosita map \citep{2024A&A...681A..77Z}. The coincident $3\arcmin\times3\arcmin$ pixel covering the region around the AGN corresponds to an X-ray flux of $F=4.3\times10^{-14}$~erg~s$^{-1}$~cm$^{-2}$ \citep{2024A&A...681A..77Z}, yielding a luminosity of $2\times10^{37}$~W ($\sim10^{44}$~erg~s$^{-1}$) in the $1-2.3$~keV energy band. This value similar to the median values of the X-ray emission detected from galaxy clusters \citep[e.g.][]{2001A&A...376..803G,2010MNRAS.407...83E,2011JApA...32..519C,2015A&A...575A.127P}. To confirm or reject the clustering hypothesis, accurate distance determinations would be needed via spectroscopic redshift measurements. Confirming that the DESI sources are indeed in a common cluster would also reveal J0843+1814 as a radio-loud brightest cluster galaxy \citep[e.g.][]{2015MNRAS.453.1201H,2015A&A...581A..23K,2021A&A...649A..42C}.

\section{Summary and conclusions}\label{sec:summary}

To find new in-beam phase-reference calibrator sources to aid VLBI measurements in the interplanetary travel phase of ESA's JUICE spacecraft, we performed nodding-style phase referencing observations along the spacecraft's expected trajectory during its Venus flyby in August 2025. Nine faint radio sources were selected from the VLASS and the  mJIVE surveys as primary targets. Additionally, three nearby RFC calibrators were observed. This way, we linked their positions to the ICRF. 

We detected four out of the nine primary targets. Unfortunately, all of the detected in-beam calibrator candidates were found to have mJy-level flux densities in mas-scale compact components, making them unsuitable as future phase-reference calibrators. On the other hand, they could still be used to verify the detection of the spacecraft during the flyby observations.

Additionally, we performed detailed analyses on two primary targets, J0841+1826 and J0843+1814, with notable astrophysical characteristics. The resolved pc-scale structure and properties of J0841+1826 can be interpreted as either a young compact symmetric object or a core--jet in a low-luminosity AGN. The radio and optical data on J0843+1814 point to a young FRII radio galaxy, with the possibility that the object belongs to nearby galaxy cluster at $z\approx0.8$.

\section*{Acknowledgements}
\small
This work was supported by the ESA PRODEX project PEA 4000136207 and the HUN-REN Hungarian Research Network.

The EVN is a joint facility of independent European, African, Asian, and North American radio astronomy institutes. Scientific results from data presented in this publication are derived from the EVN project EP129. The National Radio Astronomy Observatory is a facility of the National Science Foundation operated under cooperative agreement by Associated Universities, Inc.

This research has made use of the CIRADA cutout service at URL cutouts.cirada.ca, operated by the Canadian Initiative for Radio Astronomy Data Analysis (CIRADA). CIRADA is funded by a grant from the Canada Foundation for Innovation 2017 Innovation Fund (Project 35999), as well as by the Provinces of Ontario, British Columbia, Alberta, Manitoba and Quebec, in collaboration with the National Research Council of Canada, the US National Radio Astronomy Observatory and Australia’s Commonwealth Scientific and Industrial Research Organisation.

The Legacy Surveys consist of three individual and complementary projects: the Dark Energy Camera Legacy Survey (DECaLS; Proposal ID \#2014B-0404; PIs: David Schlegel and Arjun Dey), the Beijing-Arizona Sky Survey (BASS; NOAO Prop. ID \#2015A-0801; PIs: Zhou Xu and Xiaohui Fan), and the Mayall z-band Legacy Survey (MzLS; Prop. ID \#2016A-0453; PI: Arjun Dey). DECaLS, BASS and MzLS together include data obtained, respectively, at the Blanco telescope, Cerro Tololo Inter-American Observatory, NSF’s NOIRLab; the Bok telescope, Steward Observatory, University of Arizona; and the Mayall telescope, Kitt Peak National Observatory, NOIRLab. Pipeline processing and analyses of the data were supported by NOIRLab and the Lawrence Berkeley National Laboratory (LBNL). The Legacy Surveys project is honored to be permitted to conduct astronomical research on Iolkam Du’ag (Kitt Peak), a mountain with particular significance to the Tohono O’odham Nation.

NOIRLab is operated by the Association of Universities for Research in Astronomy (AURA) under a cooperative agreement with the National Science Foundation. LBNL is managed by the Regents of the University of California under contract to the U.S. Department of Energy.

This project used data obtained with the Dark Energy Camera (DECam), which was constructed by the Dark Energy Survey (DES) collaboration. Funding for the DES Projects has been provided by the U.S. Department of Energy, the U.S. National Science Foundation, the Ministry of Science and Education of Spain, the Science and Technology Facilities Council of the United Kingdom, the Higher Education Funding Council for England, the National Center for Supercomputing Applications at the University of Illinois at Urbana-Champaign, the Kavli Institute of Cosmological Physics at the University of Chicago, Center for Cosmology and Astro-Particle Physics at the Ohio State University, the Mitchell Institute for Fundamental Physics and Astronomy at Texas A\&M University, Financiadora de Estudos e Projetos, Fundacao Carlos Chagas Filho de Amparo, Financiadora de Estudos e Projetos, Fundacao Carlos Chagas Filho de Amparo a Pesquisa do Estado do Rio de Janeiro, Conselho Nacional de Desenvolvimento Cientifico e Tecnologico and the Ministerio da Ciencia, Tecnologia e Inovacao, the Deutsche Forschungsgemeinschaft and the Collaborating Institutions in the Dark Energy Survey. The Collaborating Institutions are Argonne National Laboratory, the University of California at Santa Cruz, the University of Cambridge, Centro de Investigaciones Energeticas, Medioambientales y Tecnologicas-Madrid, the University of Chicago, University College London, the DES-Brazil Consortium, the University of Edinburgh, the Eidgenossische Technische Hochschule (ETH) Zurich, Fermi National Accelerator Laboratory, the University of Illinois at Urbana-Champaign, the Institut de Ciencies de l’Espai (IEEC/CSIC), the Institut de Fisica d’Altes Energies, Lawrence Berkeley National Laboratory, the Ludwig Maximilians Universitat Munchen and the associated Excellence Cluster Universe, the University of Michigan, NSF’s NOIRLab, the University of Nottingham, the Ohio State University, the University of Pennsylvania, the University of Portsmouth, SLAC National Accelerator Laboratory, Stanford University, the University of Sussex, and Texas A\&M University.

BASS is a key project of the Telescope Access Program (TAP), which has been funded by the National Astronomical Observatories of China, the Chinese Academy of Sciences (the Strategic Priority Research Program ''The Emergence of Cosmological Structures'' Grant \#XDB09000000), and the Special Fund for Astronomy from the Ministry of Finance. The BASS is also supported by the External Cooperation Program of Chinese Academy of Sciences (Grant \#114A11KYSB20160057), and Chinese National Natural Science Foundation (Grant \#12120101003, \#11433005).

The Legacy Survey team makes use of data products from the Near-Earth Object Wide-field Infrared Survey Explorer (NEOWISE), which is a project of the Jet Propulsion Laboratory/California Institute of Technology. NEOWISE is funded by the National Aeronautics and Space Administration.

The Legacy Surveys imaging of the DESI footprint is supported by the Director, Office of Science, Office of High Energy Physics of the U.S. Department of Energy under Contract No. DE-AC02-05CH1123, by the National Energy Research Scientific Computing Center, a DOE Office of Science User Facility under the same contract; and by the U.S. National Science Foundation, Division of Astronomical Sciences under Contract No. AST-0950945 to NOAO.

Funding for the Sloan Digital Sky Survey V has been provided by the Alfred P. Sloan Foundation, the Heising-Simons Foundation, the National Science Foundation, and the Participating Institutions. SDSS acknowledges support and resources from the Center for High-Performance Computing at the University of Utah. SDSS telescopes are located at Apache Point Observatory, funded by the Astrophysical Research Consortium and operated by New Mexico State University, and at Las Campanas Observatory, operated by the Carnegie Institution for Science. The SDSS web site is \url{www.sdss.org}.

SDSS is managed by the Astrophysical Research Consortium for the Participating Institutions of the SDSS Collaboration, including the Carnegie Institution for Science, Chilean National Time Allocation Committee (CNTAC) ratified researchers, Caltech, the Gotham Participation Group, Harvard University, Heidelberg University, The Flatiron Institute, The Johns Hopkins University, L'Ecole polytechnique f\'{e}d\'{e}rale de Lausanne (EPFL), Leibniz-Institut f\"{u}r Astrophysik Potsdam (AIP), Max-Planck-Institut f\"{u}r Astronomie (MPIA Heidelberg), Max-Planck-Institut f\"{u}r Extraterrestrische Physik (MPE), Nanjing University, National Astronomical Observatories of China (NAOC), New Mexico State University, The Ohio State University, Pennsylvania State University, Smithsonian Astrophysical Observatory, Space Telescope Science Institute (STScI), the Stellar Astrophysics Participation Group, Universidad Nacional Aut\'{o}noma de M\'{e}xico, University of Arizona, University of Colorado Boulder, University of Illinois at Urbana-Champaign, University of Toronto, University of Utah, University of Virginia, Yale University, and Yunnan University.  
\normalsize

\begin{sidewaystable*}[ht]
\setlength{\tabcolsep}{3.3pt}\centering\small
\caption{Model parameters for the ICC sources.}\label{tab:model_target}
 \begin{tabular}{cccccccccccccc}\ttrule
Name	&	D	&	R.A.	&	Dec	&	C	&	$S_1$	&	$\vartheta_1$	&	$S_2$	&	$\vartheta_2$	&	$R_2$	&	$\phi_2$	&	$\alpha_1$	&	$\alpha_2$	\\
&&&&& (mJy) & (mas) & (mJy) & (mas) & (mas) & ($\degr$) & & \\\midrule

J0822+1930	&	n	&	08 22 44.88	&	+19 30 43.4	&		&	$<1.1$	&		&	$<1.0$	&		&		&		&		&		\\
J0828+1908	&	n	&	08 28 39.43	&	+19 08 40.6	&		&	$<1.0$	&		&	$<1.0$	&		&		&		&		&		\\
J0835+1844A	&	n	&	08 35 17.24	&	+18 44 51.0	&		&	$<1.0$	&		&	$<0.7$	&		&		&		&		&		\\
J0835+1844B	&	n	&	08 35 52.39	&	+18 44 10.0	&		&	$<1.0$	&		&	$<0.9$	&		&		&		&		&		\\
J0841+1821	&	y	&	08 41 42.310780 (0.16)	&	+18 21 56.77904 (0.30)	&	c	&	$1.6\pm0.4$	&	$1.06\pm0.19$	&	$1.6\pm0.3$	&	$0.45\pm0.07$	&		&		&	$-0.5\pm0.1$	&	$-0.5\pm0.1$	\\
J0841+1826	&	y	&	08 41 01.648311 (0.09)	&	+18 26 02.42525 (0.27)	&	j1$ ^\dagger$	&	$1.8\pm0.5$	&	$ <0.40$	&	$1.2\pm0.2$	&	$<0.16$	&	0	&	0	&	$0.0\pm0.2$	&	$-0.1\pm0.2$	\\
	&$\phantom{	y	}$&		&		&	j2	&	--	&	--	&	$0.8\pm0.2$	&	$0.29\pm0.10$	&	$1.5\pm0.1$	&	$45\pm2$	&		&		\\
J0843+1814	&	y	&	08 43 08.583913 (0.22)	&	+18 14 00.37865 (0.33)	&	c	&	$0.8\pm0.4$	&	$ <0.89$	&	$0.7\pm0.2$	&	$0.27\pm0.05$	&		&		&	$-0.2\pm0.3$	&	$-0.3\pm0.2$	\\
J0843+1815	&	n	&	08 43 32.58	&	+18 15 31.1	&		&	$<1.4$	&		&	$<1.4$	&		&		&		&		&		\\
J0843+1816	&	y	&	08 43 05.383596 (0.23)	&	+18 16 05.08225 (0.34)	&	c	&	$0.5\pm0.2$	&	$ <0.76$	&	$0.5\pm0.2$	&	$0.88\pm0.24$	&		&		&	$-0.8\pm0.3$	&	$-0.7\pm0.2$	\\
\bottomrule
\end{tabular}
\begin{tabnote}
{\textit{Notes:} Col.~1 -- short name, Col.~2 -- source detection status (y = yes, n = no), Col.~3--4 -- right ascension and declination; for detected sources, mean value of the the two observations is given with the astrometric errors (mas) in parentheses, otherwise VLASS coordinates are displayed, Col.~5 -- model component; $^\dagger$ denotes the the single unresolved component, Col.~6 -- flux density or $5\sigma$ rms upper limit assuming an unresolved source (Cal1) , Col.~7 -- model size (FWHM) or upper limit corresponding to the minimum resolvable angular size following \citet{2005AJ....130.2473K} (Cal1), Col.~8 -- flux density or $5\sigma$ rms upper limit (Cal2), Col.~9 -- model size or FWHM upper limit (Cal2), Col.~10--11 -- VLBI spectral index from $1.4$~GHz mJIVE and $8.4$~GHz EVN observations for the Cal1 and Cal2 data sets, respectively. Cal1 and Cal2 denote data sets fringe-fitted with calibrators listed in Table~\ref{tab:obs}.}
\end{tabnote}

\vspace{2cm}\caption{Model parameters for the RFC sources}
 \label{tab:model_rfc}
 \hspace*{-0.65cm}\begin{tabular}{cccccccccccccccc}\ttrule 
Name			&	R.A.	&	Dec&	C	&	$S_1$	&	$\vartheta_1$	&	$R_1$	&	$\phi_1$	&	$S_2$	&	$\vartheta_2$	&	$R_2$	&	$\phi_2$	&	$S_\mathrm{SF}$	&	$\vartheta_\mathrm{SF}$	&	$R_\mathrm{SF}$	&	$\phi_\mathrm{SF}$	\\

&&&& (mJy) & (mas) & (mas) & ($\degr$) & (mJy) & (mas) & (mas) & ($\degr$) & (mJy) & (mas) & (mas) & ($\degr$)\\
\midrule

J0832+1953			&	08 32 00.158144 (0.12)	&	+19 53 12.17581 (0.18)	&	c	&	$46.6\pm5.4$	&	$<0.20$	&	0	&	0	&	$46.6\pm5.8$	&	$<0.25$	&	0	&	0	&	$56.1\pm6.3$	&	$1.11\pm0.04$	&	0	&	0	\\
J0839+1921			&	08 39 06.943947 (0.12)	&	+19 21 48.68536 (0.29)	&	c	&	$37.2\pm4.1$	&	$<0.21$	&	0	&	0	&	$37.5\pm2.4$	&	$0.44\pm0.02$	&	0	&	0	&	$43.5\pm4.5$	&	$0.32\pm0.01$	&	0	&	0	\\
			&		&		&	j	&	$2.7\pm0.4$	&	$2.73\pm0.86$	&	$6.1\pm0.4$	&	$20\pm4$	&	$2.1\pm0.3$	&	$1.43\pm0.30$	&	$4.8\pm0.2$	&	$12\pm2$	&	$2.5\pm0.4$	&	$1.86\pm0.23$	&	$4.8\pm0.1$	&	$3\pm1$	\\
J0846+1735			&	08 46 00.033774 (0.34)	&	+17 35 25.11800 (0.42)	&	c	&	$33.1\pm3.8$	&	$1.12\pm0.05$	&	0	&	0	&	$33.1\pm3.3$	&	$1.36\pm0.09$	&	0	&	0	&	$45.0\pm5.1$	&	$1.024\pm0.039$	&	0	&	0	\\
\bottomrule\end{tabular}
\begin{tabnote}
{\textit{Notes:} Col.~1 -- short name, Col.~2--3 -- mean right ascension and declination of the two observations (Cal1, Cal2) with astrometric errors (mas) in parentheses, Col.~4 -- model component, Col.~5 -- flux density (Cal1), Col.~6-- model FWHM size or upper limit (Cal1), Col.~7--8 -- model radius and position angle relative to the core (Cal1), Col.~9 -- flux density (Cal2), Col.~10 -- model FWHM size (Cal2), Col.~11--12 -- model radius and position angle relative to the core (Cal2), Col.~13 -- flux density (SF), Col.~14 -- model FWHM size (SF), Col.~15--16 -- model radius and position angle relative to the core (SF). Cal1 and Cal2 denote data sets fringe-fitted with calibrators listed in Table~\ref{tab:obs}, while SF refers to the self-fringe-fitted data.}
\end{tabnote}
\end{sidewaystable*}

\section*{Data Availability}
Raw data from the new observations can be obtained from the EVN data archive\footnote{\url{http://archive.jive.nl}} under project code EP129 (\url{https://doi.org/10.48717/52nc-p759}), and the archival VLA observations are available on the NRAO data archive\footnote{\url{https://data.nrao.edu}} under project codes AV162 and LAWR. 

\bibliographystyle{pasa-mnras.bst}
\bibliography{pridevenus.bib}

\begin{appendix}\onecolumn
\section{Calibrator sources}\label{sec:calibrators}
\begin{figure*}[h!]
 \centering
 \includegraphics[width=0.75\linewidth]{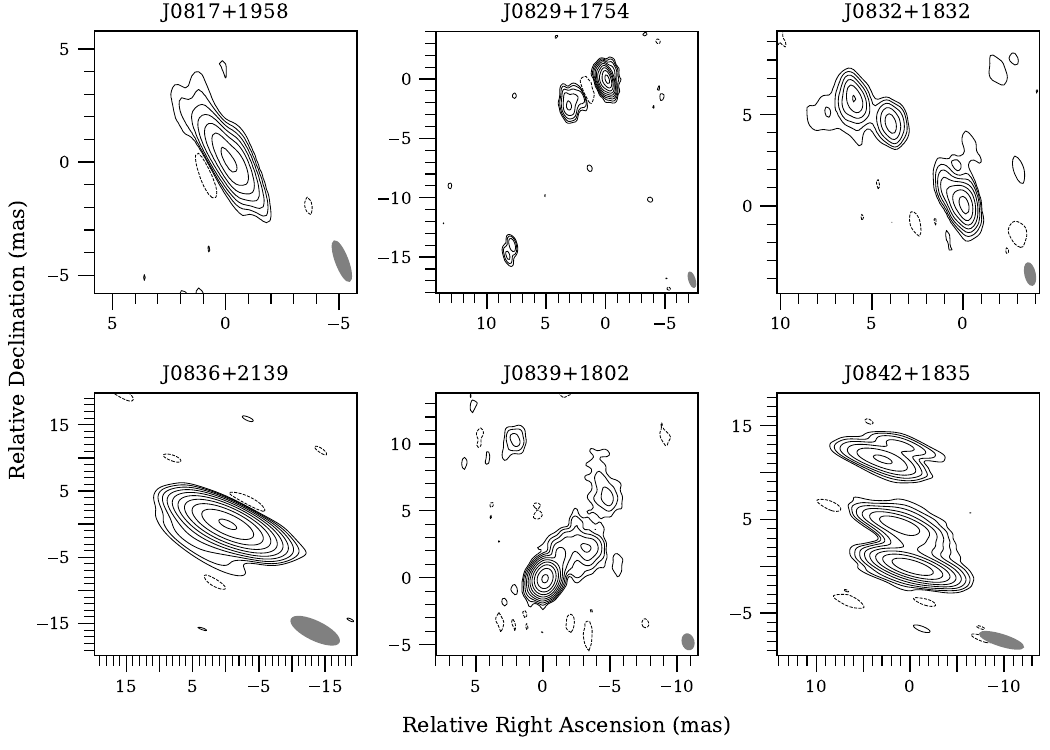}
 \caption{Clean radio maps at $8.4$~GHz of the phase-reference calibrators observed in the EP129 experiment. The first contours are drawn at $\pm3\sigma$ rms noise level. The positive levels increase by a factor of 2. The restoring beams are shown in the lower right corners. Image parameters are listed in Table~\ref{tab:cal}.}
 \label{fig:cal}
\end{figure*}

In Figure~\ref{fig:cal}, we present the clean maps of the phase-reference calibrators used in the experiment. Image parameters are listed in Table~\ref{tab:cal}. Four out of the six calibrator sources were resolved at VLBI scales with the EVN, with an extended morphology typically in the north--south direction. As discussed in Section~\ref{sec:errors}, these extended structures could affect the precision of the relative astrometry of target sources, resulting in relatively larger differences between the two declination coordinate solutions determined for the targets from experiments Cal1 and Cal2.

\begin{table}[ht]
 \centering
 \caption{Image parameters for clean maps of calibrator sources shown in Figure~\ref{fig:cal}. }
 \label{tab:cal}
 \begin{tabular}{cccccccc}
 \ttrule
Name & R.A. & Dec &$P$	&	rms	&	$b_\mathrm{min}$	&	$b_\mathrm{maj}$	&	$b_\mathrm{PA}$	\\

&&& (mJy~beam$^{-1}$) & (mJy~beam$^{-1}$) & (mas) & (mas) & ($\degr$)\\
\midrule
J0817+1958&08 17 05.493282$^*$&+19 58 42.89847$^*$&23.1&0.1&0.5&1.9&21\\
J0829+1754&08 29 04.828611~~~&+17 54 15.86399~~~&36.4&0.1&0.5&1.3&17\\
J0832+1832&08 32 16.040283~~~&+18 32 12.13313~~~&86.2&0.3&0.5&1.2&12\\
J0836+2139&08 36 16.216892~~~&+21 39 03.58055~~~&95.2&0.1&2.8&7.9&64\\
J0839+1802&08 39 30.721364~~~&+18 02 47.14274~~~&95.5&0.1&0.8&1.2&15\\
J0842+1835&08 42 05.094174~~~&+18 35 40.99051~~~&293.1&0.4&1.2&4.9&73\\\bottomrule
\end{tabular} 
\begin{tabnote}
{\textit{Notes:} Col.~1 -- short name, Col.~2--3 -- ICRF3 \citep{2020A&A...644A.159C} or ($^*$) RFC \citep{2025ApJS..276...38P} right ascension and declination coordinates (image centres), Col.~4 -- peak intensity, Col.~5 -- rms noise, Col.~6--8 -- minor and major axis FWHM size and major axis position angle of the restoring beam.}
\end{tabnote}
\end{table}

\end{appendix}

\end{document}